\documentclass[%
  reprint,
  superscriptaddress,
  amsmath,amssymb,
  aps,
  prb,
]{revtex4-1}
\pdfoutput=1
\usepackage{graphicx}% Include figure files
\usepackage{dcolumn}% Align table columns on decimal point
\usepackage{bm}% bold math
\usepackage{siunitx}
\usepackage{xcolor}
\usepackage{natbib}
\DeclareSIUnit{\belmilliwatt}{Bm}
\DeclareSIUnit{\dBm}{\deci\belmilliwatt}
\begin{document}

%\preprint{APS/123-QED}

\title{Rapid detection of coherent tunneling in an InAs nanowire quantum dot through dispersive gate sensing}
% \title{Dispersive sensing of charge-tunneling on the microsecond scale in an InAs nanowire double quantum dot}% Force line breaks with \\

\author{Damaz de Jong}
\affiliation{QuTech and Kavli Institute of Nanoscience, Delft University of Technology, 2600 GA Delft, The Netherlands}

\author{Jasper van Veen}
\affiliation{QuTech and Kavli Institute of Nanoscience, Delft University of Technology, 2600 GA Delft, The Netherlands}

\author{Luca Binci}
\affiliation{QuTech and Kavli Institute of Nanoscience, Delft University of Technology, 2600 GA Delft, The Netherlands}

\author{Amrita Singh}
\affiliation{QuTech and Kavli Institute of Nanoscience, Delft University of Technology, 2600 GA Delft, The Netherlands}

\author{Peter Krogstrup}
\affiliation{Center for Quantum Devices and Microsoft Quantum Materials Lab, Niels Bohr Institute, University of Copenhagen,
Copenhagen, Denmark}

\author{Leo P. Kouwenhoven}
\affiliation{QuTech and Kavli Institute of Nanoscience, Delft University of Technology, 2600 GA Delft, The Netherlands}
\affiliation{Microsoft Quantum Lab Delft, Delft University of Technology, 2600 GA Delft, The Netherlands}
\author{Wolfgang Pfaff}
\email{wolfgang.pfaff@microsoft.com}
\affiliation{Microsoft Quantum Lab Delft, Delft University of Technology, 2600 GA Delft, The Netherlands}
\author{John D. Watson}
\affiliation{Microsoft Quantum Lab Delft, Delft University of Technology, 2600 GA Delft, The Netherlands}

\date{\today}% It is always \today, today,

\begin{abstract}
Dispersive sensing is a powerful technique that enables scalable and high-fidelity readout of solid-state quantum bits.
In particular, gate-based dispersive sensing has been proposed as the readout mechanism for future topological qubits, which can be measured by single electrons tunneling through zero-energy modes.
The development of such a readout requires resolving the coherent charge tunneling amplitude from a quantum dot in a Majorana-zero-mode host system faithfully on short time scales.
Here, we demonstrate rapid single-shot detection of a coherent single-electron tunneling amplitude between InAs nanowire quantum dots.
We have realized a sensitive dispersive detection circuit by connecting a sub-GHz, lumped element microwave resonator to a high-lever arm gate on one of dots.
The resulting large dot-resonator coupling leads to an observed dispersive shift that is of the order of the resonator linewidth at charge degeneracy.
This shift enables us to differentiate between Coulomb blockade and resonance -– corresponding to the scenarios expected for qubit state readout -– with a signal to noise ratio exceeding 2 for an integration time of \SI{1}{\micro \second}.
Our result paves the way for single shot measurements of fermion parity on microsecond timescales in topological qubits.
 
\end{abstract}

\maketitle

\section{\label{sec:introduction}Introduction}
\begin{figure}
	\includegraphics{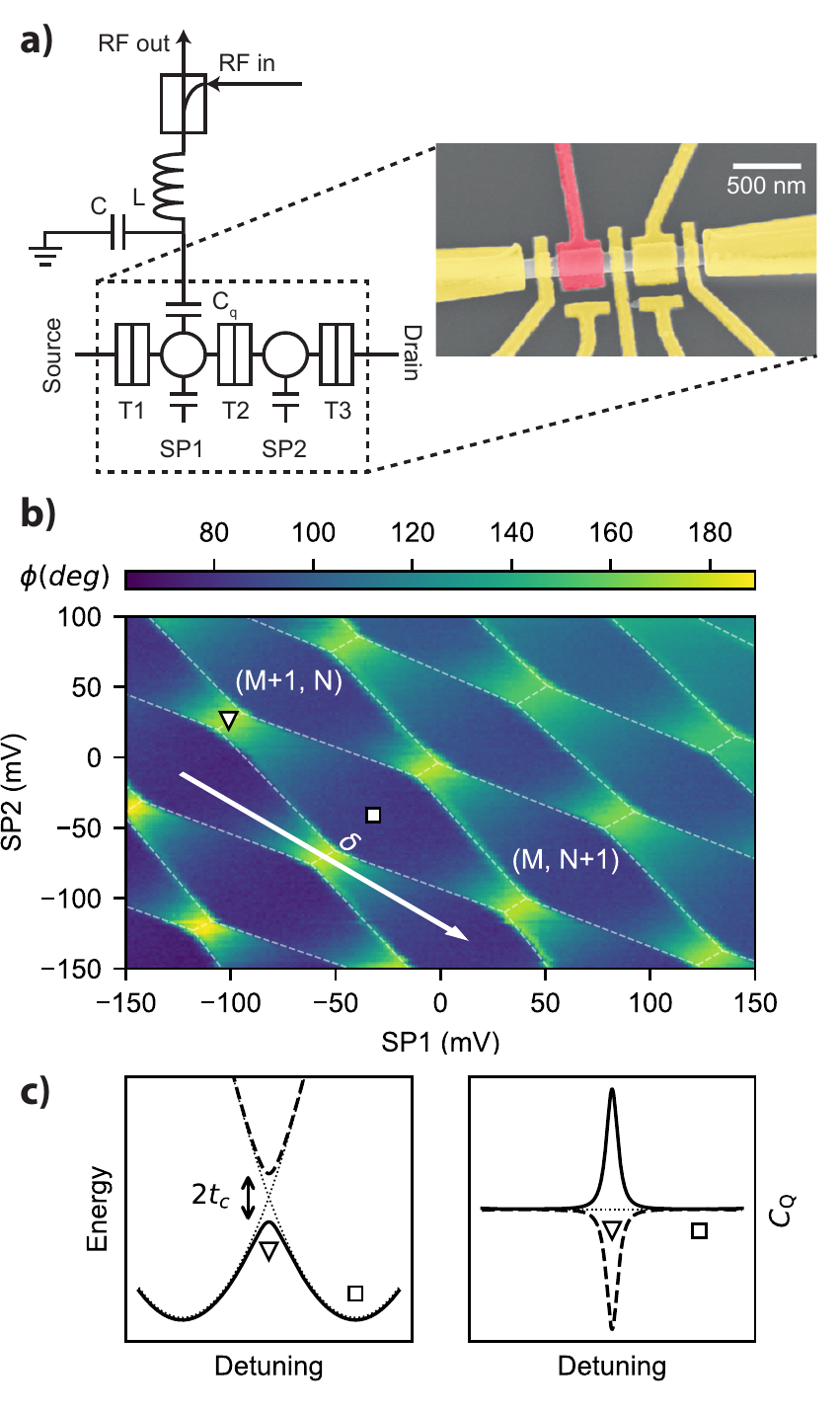}
	\caption{
	{\bf Dispersive sensing on an InAs nanowire double quantum dot.}
	{\bf a)} Schematic of the of experiment measurement setup. One of the quantum dots is capacitively coupled to a resonant circuit that is probed in reflectometry. Inset: False-colored electron micrograph of a nominally identical device. The sensing top gate is colored red. {\bf b)} Charge stability diagram measured with the gate resonator. The dashed lines are guides to the eye. The triangle marker denotes charge degeneracy while the square marker denotes Coulomb blockade. The arrow denotes the detuning from charge degeneracy, $\delta$. {\bf c)} Sketch of the energy levels and resulting quantum capacitance vs.\ detuning. Solid lines: ground state; dashed lines: first excited state; dotted line: case of no interdot tunneling.}
\end{figure}

Dispersive sensing is a promising measurement technique that enables high-fidelity readout of solid state quantum bits, such as superconducting qubits \cite{Blais2004, Wallraff2005} or spins\cite{Betz2015}.
Recently, dispersive readout has also been proposed for future topological qubits based on Majorana zero modes (MZMs) \cite{Plugge2017, Karzig2017}.
In particular, gate-based dispersive readout can be used to measure an electron tunneling rate in the system which in turn reflects the state of the qubit \cite{Colless2013}.
As a result of this difference in tunnel coupling, different qubit states can impart a different dispersive shift on a resonator coupled to the gate electrode.
This frequency shift can be probed on very fast time scales, using state-of-the-art radio frequency (RF) techniques, and in a quantum non-demolition manner with minimal perturbation \cite{Blais2004, Vijay2011}.

High-fidelity, quantum non-demolition measurements require fast readout with high signal-to-noise ratio (SNR).
This is particularly crucial for measurement-based quantum computation, such as proposed for MZM-based architectures \cite{Bonderson2008, Plugge2017, Karzig2017}.
So far, however, the frequency shift of dispersive gate sensors has been fairly small, on the order of a degree\cite{Lambert2016, Betz2015, Colless2013, Frey2012, Schroer2012}; 
correspondingly, the required readout times to resolve a difference in tunnel coupling has been in the range of milliseconds \cite{West2018, Pakkiam2018, Urdampilleta2018}.
It is thus of great interest to find avenues toward increasing the attainable SNR, and achieve readout on the sub-microsecond scale, as available for other solid-state qubit platforms\cite{Walter2017}.
%\marginpar{\todo{walraff$\approx$10us, need more}}.

In this letter we show rapid dispersive sensing in an InAs nanowire double quantum dot system.
InAs nanowires have been studied in the context of spin qubits\cite{Nadj-Perge2010, Petersson2012}, but have recently gained also significant attention as host system for MZMs that could enable the realization of topological qubits\cite{Das2012, Albrecht2016}.
We demonstrate a sensitive gate sensor based on a large-lever arm top-gate that is connected to an off-chip, lumped element resonant circuit probed with reflectometry\cite{Hornibrook2014}.
In particular, we show a dispersive shift close to \SI{1}{\mega \hertz} which is on the order of the linewidth of the resonator; this results in a detected phase shift that approaches the maximally possible value of $\pi$.
We study in detail the magnitude of the dispersive shift both as a function of tunnel coupling and readout power; we find, in agreement with theory, that the attainable shift is ultimately set by the magnitude of the tunneling rate and the resonator frequency.
The large shift allows us to resolve a difference in tunneling rate with an SNR of up to 2 within \SI{1}{\micro \second}.

\section{Experimental approach and setup}

The coherent tunneling amplitude $t_\mathrm{C}$, between two single-particle levels in weakly coupled quantum dots can be detected through an arising change in differential capacitance\cite{Ota2010,Ashoori1992}.
The coupling affects the expectation value of charge on either island.
Since level detuning and coupling is influenced by external gate voltage, the dependence of induced charge on gate voltage, i.e., the differential capacitance $C = \partial Q / \partial V_\mathrm{g}$, depends on the coupling.
This effect can be described within the framework of circuit quantum electrodynamics (circuit QED)\cite{Blais2004} or as a `quantum capacitance'\cite{Duty2005} and measured by monitoring the change in differential capacitance through an external tank circuit.
Our aim is to determine how fast the tunneling amplitude can be detected; this maps to the projected readout performance for MZM qubits\cite{Plugge2017, Karzig2017} where the magnitude of the tunnel coupling is the qubit readout signal.

Our experiment approach is schematically depicted in Fig.~1a. 
We form two quantum dots in an InAs nanowire where the interdot coupling can be set through a gate voltage.
We designate one of the dots as the `sensor', whereas the other dot is merely used as an auxiliary single level system, in lieu of MZMs.
To achieve a large signal from the interdot coupling, we connect a gate with a large lever arm to a resonant circuit.
The goal of the experiment is then to resolve a change in resonance frequency,
\begin{equation}
	\label{eq:cq-frqshift}
	\delta\omega = \sqrt{LC}^{-1} - \sqrt{L(C+C_\mathrm{q})}^{-1},
\end{equation}
that arises from the tunneling-dependent quantum capacitance $C_\mathrm{q}$ (Fig.~1c).

To realize this experiment we have fabricated a double quantum dot in an InAs nanowire which was deposited on an intrinsic silicon substrate with a \SI{20}{\nano \meter} SiNx dielectric layer deposited with LPCVD after removing the native SiO2.
A \SI{10}{\nano \meter} AlOx dielectric layer is deposited using atomic layer deposition (ALD) between the nanowire and the top gates, which ensures a large lever arm from the gates to the underlying quantum dots.
A false color SEM image of a similar device is shown in Fig.~1a.
Using top gates T1, T2, and T3, a double dot is defined in the nanowire by pinching off the coupling to the leads and between the two dots.
The top gate of the sensing dot is wire-bonded to a lumped-element resonator that was fabricated on a separate chip\cite{Hornibrook2014}.
The sample is cooled down in a dilution refrigerator with a base temperature of \SI{20}{\milli \kelvin}.
This resonator response is then probed using standard RF heterodyne techniques (Fig.~1a).

\section{\label{sec:results}Results}

\subsection{Observation of quantum capacitance and dispersive shift}

We begin by characterizing the change in resonator response resulting from coherent tunneling between the two quantum dots.
To this end we first tune the device to a regime where the dot charge states strongly hybridize on resonance.
We then record the phase response of a reflected probe field as a function of the two plunger gates, SP1 and SP2. (Fig.~1b).
The resulting charge stability diagram shows a prominent phase shift at charge degeneracy, hinting at a large dispersive shift of the resonator frequency.
We attribute the substantial magnitude of the observed phase shift in this regime to the large lever arm of the sensing gate\cite{Blais2004, Duty2005}.
From independent Coulomb blockade measurements we estimate this to be $\alpha = C_\mathrm{g}/C_\Sigma \approx 0.75$, where $C_\mathrm{g}$ is the capacitance of the gate to the sensing dot, and $C_\Sigma$ is the total capacitance seen by the dot.

\begin{figure}
	\includegraphics{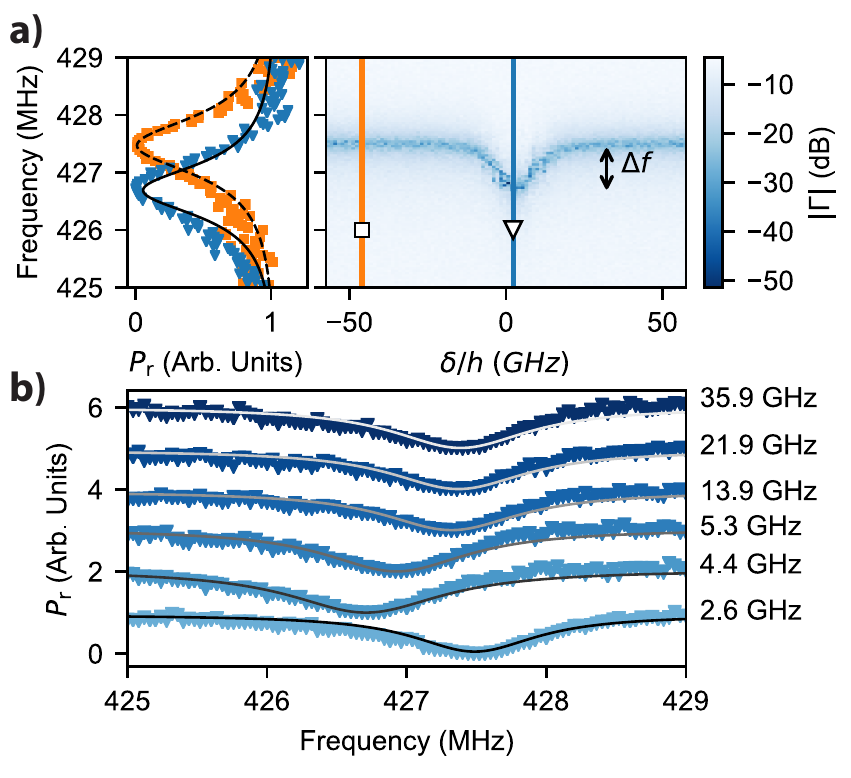}
	\caption{
	{\bf Charge-resonator coupling.}	
	{\bf a)} Right panel: Resonator reflection spectrum measured from the difference between injected ($P_\mathrm{RF}$) and reflected RF power ($P_\mathrm{r}$) corrected for estimated attenuation and gain in the setup, as a function of detuning $\delta$. T2-gate voltage was \SI{-0.768}{\volt} for this data. 
	Left panel: Line cuts in blockade (orange; square) and at degeneracy (blue; triangle) together with fits (black) to Eq.~\eqref{eq:aout}. 
	{\bf b)} Resonator spectroscopy at charge degeneracy for different tunneling rates together with the fit to Eq.~\eqref{eq:aout}. Traces are offset for clarity. Tunnel rates $t_\mathrm C / h$ extracted from the fit are indicated on the right.}
\end{figure}

The relation between the dispersive shift and the magnitude of the interdot coupling lies at the heart of the $C_\mathrm{q}$ detection scheme; we therefore focus next on modeling this relation from our data following earlier work performed on semiconductor dots in cQED environments\cite{Petersson2012, Frey2012}.
Near charge degeneracy the eigenstates of the double dot are superpositions of a charge delocalized between the two dots, with energy splitting $\Omega = \sqrt{4t_\mathrm{C}^2 + \delta^2}$, where $t_\mathrm C$ is the tunnel coupling, and $\delta$ is the detuning of the two dots (Fig.~1c)\cite{Wiel2002}. 
To determine the tunnel coupling, we measure the resonator response as a function of $\delta$ and the detuning of the drive from the resonance frequency (Fig.~2a).
The reflected probe signal can be developed in a cQED approach from the input-output relations\cite{Blais2004, Petersson2012},
\begin{equation}
	\label{eq:aout}
	\frac{a_\mathrm{out}}{a_\mathrm{in}} = 1 +
		\frac{\mathrm{i} \kappa_\mathrm{ext}}
			{-\mathrm{i}\kappa/2 + \Delta\omega + g\chi} .
\end{equation}
Hereby, $a_{\mathrm{in,out}}$ are the complex input and output signals; 
$\kappa = \kappa_\mathrm{int} + \kappa_\mathrm{ext}$ is the total resonator damping rate, composed of internal losses $\kappa_\mathrm{int}$ and external coupling $\kappa_\mathrm{ext}$; 
$\Delta\omega$ is the detuning of the drive from resonance; $g = g_0 (2 t_\mathrm{C}/\Omega)$ is the effective coupling strength with $g_0$ being the Jaynes-Cummings coupling; 
and $\chi$ is the susceptibility of the double quantum dot that depends on the dephasing rate $\gamma$ and detuning between charge dipole and resonator,
\begin{equation}
	\chi = g / (\omega_0 - \Omega + \mathrm{i}\gamma/2).
\end{equation}

Figure~2a shows the evolution of the dispersive shift as we tune the double dot between Coulomb blockade regime and charge degeneracy, for one particular tunnel gate setting.
Fitting this data yields the tunnel coupling, as well as the relevant parameters characterizing circuit and resonator-dot coupling.
In particular, we find $Q = \omega_0/\kappa \approx 350$, and $g_0/2\pi \approx 60\,\mathrm{MHz}$, consistent with the large lever arm.
This procedure allows us now to correlate the tunnel coupling and the dispersive shift with the gate voltage on electrode T2 (Fig.~2b).

\begin{figure}
	\includegraphics{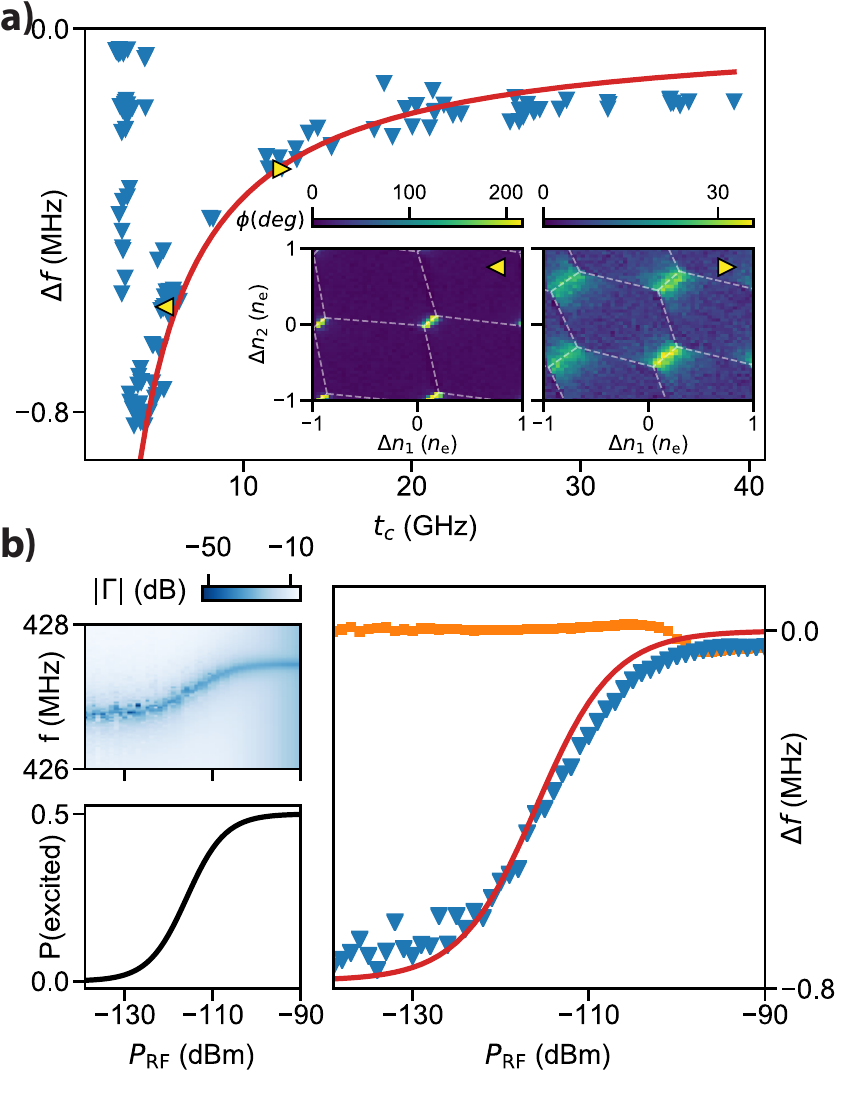}
	\caption{
	{\bf Evolution and modeling of the dispersive shift.}	
	{\bf a)} Frequency shift as a function of tunneling rate, extracted from fitting spectroscopy data to Eq.~\eqref{eq:aout}. Solid line: independent theoretical prediction from Eq.~\eqref{eq:cqtc}. Inset: charge stability diagrams for tunneling rates corresponding to the yellow markers.
	{\bf b)} Resonator response as a function of frequency and power. Power is given at the sample level; this is attenuated by a total of $\sim 79$\,dB after the generator.
	Top left panel: resonator spectroscopy as function of RF power.
	Bottom left panel: calculated steady state population in the excited state.
	Right panel: Resonator shift in blockade (orange), and on degeneracy (blue). Red: prediction from the excited state population by assuming that the net shift is given by the population-weighted average between ground and excited state shifts.}
\end{figure}

\subsection{Quantitative model of the dispersive shift}

Having established the means to analyze the resonator response, we now investigate the change in resonator frequency as a function of double dot properties.
Figure~3a shows the magnitude of the dispersive shift at charge degeneracy as a function of tunnel coupling.
This shift can be predicted using the quantum capacitance picture; from determining the expectation value of charge on the sensing dot one expects\cite{Duty2005,Petersson2010}
\begin{equation}
	\label{eq:cqtc}
	C_\mathrm{q} = \frac{\alpha^2 e^2}{4 t_\mathrm{C}},
\end{equation}
where $e$ is the electron charge;
this relation straight-forwardly yields the frequency shift through Eq.~(\ref{eq:cq-frqshift}).
We find that this prediction agrees well with our data for tunnel couplings $t_\mathrm{C}/h \gtrsim 4\,\mathrm{GHz}$.
The effect of reduced frequency shift with increasing tunnel coupling is reflected also in the familiar geometry of charge stability diagrams (Fig.~3a, inset).
For small tunnel couplings we observe a reduction in the shift; this behavior is likely due to noise in the system, such as thermal fluctuations\cite{Schroer2012} or charge fluctuations on the gates (i.e., fluctuations in $\delta$).
%\marginpar{\todo{cite something?}}; 
This noise would effectively blur out the $C_\mathrm{q}$ peak as it narrows with decreasing $t_\mathrm C$.

A natural question that arises is in which regimes this simple description holds.
In particular, from the quantum capacitance picture one could naively expect that it is always possible to increase the power of the readout tone to increase the signal-to-noise ratio (SNR).
However, this view ignores any internal dynamics of the quantum dot system that can impact the dispersive shift.
Most importantly, increasing the AC voltage of the readout drive can induce transitions of the ground state to the excited state of the double dot, resulting in an incoherent mixture.
Since the dispersive shift from the excited state is opposite to that of the ground state, excitation would thus lead to a reduction of the measured shift.

In Figure 3b we show the evolution of the dispersive shift when increasing the readout drive amplitude; indeed, the shift disappears entirely at large drive amplitudes.
We compare this data to a model in which we compute the excitation of the double dot by assuming that the readout drive acts as a detuned Rabi drive (with detuning $\omega_0 - 2t_\mathrm C / h$) and the double dot dephases quickly.
We find that the double dot approaches a fully mixed state in the same range in which the disappearance of the shift occurs; 
the resulting predicted dispersive shift is in very good agreement with the data.
% We note that this effect is in close relation to the dissappearance of the dispersive shift in superconducting quantum circuits [Reed, Bishop], albeit with a very quickly dephasing two-level system.

We can therefore conclude that the tunnel coupling has two competing influences on the observed resonator shift:
For one, the shift gets larger for decreasing $t_\mathrm C$ Eq.~(\ref{eq:cqtc}).
On the other hand, in the present setup a decreased tunnel coupling results in reduced drive detuning; this in turn increases excited state population, reducing the shift again.

\subsection{SNR for detecting a tunnel amplitude}
\begin{figure}
	\includegraphics{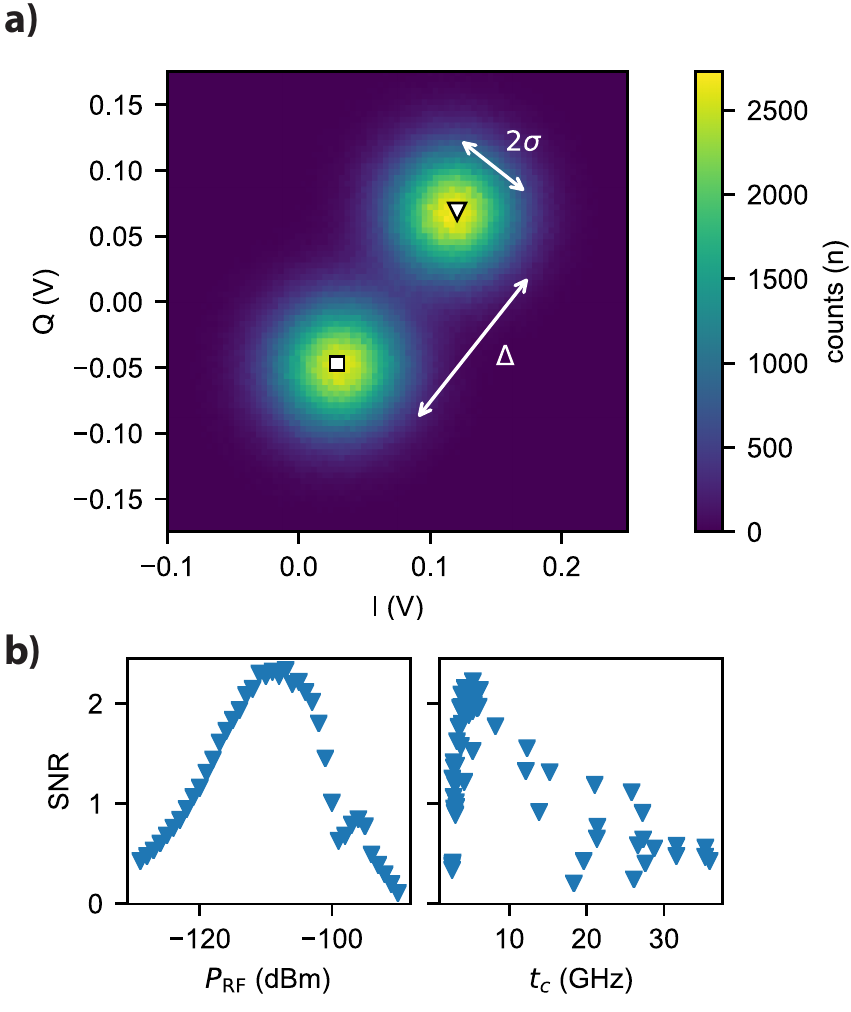}
	\caption {{\bf Readout SNR.}
	{\bf a)} Histogram of resonator reflection measurements in Coulomb blockade and charge degeneracy.
	This data was taken with a tunnel coupling of \SI{4.3}{\giga \hertz} and a readout power of $P_\mathrm{RF}=\SI{-105}{\dBm}$.
	Each count corresponds to an integration time of \SI{1}{\micro \second}.
	The SNR is defined as $\Delta / 2 \sigma$.
	{\bf b)} Attained SNR as a function of readout power (left panel) and tunnel coupling (right panel).}
\end{figure}

In order to show the feasibility of dispersive gate sensing for qubit readout, we finally investigate the time-resolved resonator response.
In particular, we aim to show that the difference in charge hybridization between Coulomb blockade and charge degeneracy can be obtained on fast time scales\cite{Plugge2017, Karzig2017}
To do so, we repeatedly measure the RF-signal in Coulomb blockade and on charge degeneracy by switching between the two points in the charge stability diagram.
The signal is then binned in \SI{1}{\micro \second} intervals and for each interval, the in-phase and quadrature components of the signal are extracted and represented in a histogram (Fig.~4a).
From Gaussian fits we can then extract the SNR, which is given by the distance between the two distributions, $\Delta$, divided by their full width, $2\sigma$.
These widths are set by the noise in the system, which is dominated by the thermal contribution of the cryogenic amplifier. 
From independent measurements we estimate the equivalent noise temperature of the readout circuit to be around \SI{4}{\kelvin}.

In Figure~4b we show the attained SNR per \SI{1}{\micro \second} `shot' as a function of readout power and tunnel coupling. The SNR reaches its peak value of $> 2$ for an RF power of $P_\mathrm{RF} \approx -109\,\mathrm{dBm}$ and a tunnel coupling of $\sim 5\,\mathrm{GHz}$. Since the signal is largely set by the frequency shift, the dependence of the SNR on $t_\mathrm{C}$ closely follows the evolution of the dispersive shift shown in Fig.~3a. The power dependence results from the competition between double dot excitation and signal increase. The optimal power is reached at the point where the diminishing frequency shift starts dominating over the improvement gained from larger accuracy in the estimation of $I$ and $Q$.

%{\bf Here we need a proper treatment of why the SNR is what it is. What's the noise temperature? How does it relate to SNR? ... At the moment there is %no good explanation of the magnitude of the SNR.}

% The SNR inversely proportional to the square root of the integration time.
% Therefore the acquisition time is fixed to $\SI{1}{\micro\second}$ to ensure proper comparison.
% %The SNR also depends on the measurement frequency, the dependence is shown in the inset of Fig~4c.
% As our noise is dominated by the cryogenic amplifier, increasing the power increases the SNR because the relative noise contribution decreases.
% For higher power the signal starts to decrease however, until it disappears, see Fig. 3c.
% These two effects compete for an optimal SNR around \SI{-30}{dBm} as shown in the inset of Fig 4b.
% Finally, the SNR depends on the readout frequency which can be optimized by sweeping over frequency.

% Since the signal depends on the value of the tunnelcoupling, the SNR is also expected to depend on $t_\mathrm{c}$, which is tuned by T2.
% For each value of T2, we extract the value of $t_\mathrm{c}$ similar to Fig.~3a. The readout frequency and power are subsequently optimized independently for every measurement and the SNR is measured (Fig.4b).
% From this measurement we conclude that the tunnelcoupling should either be optimized during fabrication, or needs to be tunable to allow for in situ optimization of the tunnelcoupling.

\section{\label{sec:conclusions}Conclusions}

We have performed gate-based dispersive sensing on a double quantum dot in an InAs nanowire.
The observed charge-tunneling induced dispersive shift on our resonator is comparable to the resonator linewidth, enabling fast detection of the presence of the tunnel amplitude with high SNR.
Notably, this result was achieved with a low-Q, lumped-element resonator operating at a frequency of less than \SI{1}{\giga \hertz}; 
these types of resonators hold great promise for scalable readout due to their reduced footprint compared to high-Q, CPW resonators that are more traditionally used in cQED\cite{Blais2004}.
Utilizing the large resonator shift, we have shown that states corresponding to different charge hybridizations can be distinguished in $\SI{1}{\micro\second}$ measurements while retaining an SNR exceeding two in our experimental setup.
We have further established that the factor that predominantly limits the SNR is the tunnel coupling. 
Its magnitude determines the dispersive shift, and its detuning from the resonator frequency places a limit on the readout power that can be used before adverse effects take over.

%Our results show that 1us readout of NW based qubits can be done.
Our results show that high-fidelity measurements of semiconductor nanowire-based qubits could be performed using gate-sensing on the single-microsecond scale. 
This is particularly promising for MZM-based topological qubits that could be realized in nanowire networks\cite{Plugge2017, Karzig2017}. 
Since our work illustrates the dominating factor of only a few key device parameters --- such as electron tunneling rate, gate lever arm, and resonator frequency --- our results can provide important guidance for the design of qubit and measurement circuits.
We further expect that existing technology could be used to lower the noise temperature of the cryogenic amplifier\cite{Castellanos-Beltran2008, Macklin2015,Stehlik2015,Schupp2018} or optimize the sensing circuits\cite{Ahmed2018} in order to enhance the attainable SNR further, and reduce the required measurement time.

\section{\label{sec:acknowledgements}Acknowledgements}

We thank J.K.~Gamble, T.~Karzig, R.M.~Lutchyn, and K.~van Hoogdalem for useful discussions.
We further thank O.W.B.~Benningshof, J.D.~Mensingh, R.N.~Schouten, M.J.~Tiggelman, and R.F.L.~Vermeulen for valuable technical assistance; R.~McNeil for help with fabrication; and J.M.~Hornibrook and D.J.~Reilly for providing the frequency multiplexing chips.
This work has been supported by the Netherlands Organization for Scientific Research (NWO), Microsoft, the Danish National Research Foundation, and the European Research Council.

\end{document}